\documentclass[rnote]{aa} 
\usepackage{graphicx}
\usepackage{txfonts}
\usepackage{epsfig}
\def\msole {~M_{\odot}}
\begin{document}
   \title{A search for evidence of irradiation in Centaurus X-4 during
quiescence\thanks{The results reported in this paper are partially based on 
observations carried out at ESO, La Silla, Chile (67.D-0116).}}


   \author{P. D'Avanzo\inst{1,2} 
       \and
       T. Mu\~noz-Darias\inst{3} 
       \and
       J. Casares\inst{3}
       \and
       I. G. Mart\'inez-Pais\inst{3} 
       \and
       S.~Campana\inst{1}
          }

   \offprints{Paolo D'Avanzo - \email{paolo.davanzo@brera.inaf.it}}

   \institute{INAF-Osservatorio Astronomico di Brera, Via Bianchi 46, I--23807
Merate (Lc), Italy
\and
Universit\`a degli Studi dell'Insubria, Dipartimento di Fisica e Matematica, Via Valleggio
11, I--22100 Como, Italy
\and
Instituto de Astrof\`{\i}sica de Canarias, 38200 La Laguna,
Tenerife, Spain
   }

   \date{Received; accepted}

 
  \abstract
      {}
{We present a study of the neutron star X-Ray Transient Cen X-4. 
Our aim is to look for any evidence of irradiation of the companion with a
detailed analysis of its radial velocity curve, relative contribution of the donor star and Doppler
tomography of the main emission lines.}
{To improve our study all our data are compared with a set of 
simulations that consider different physical parameters of the system, like the disc aperture angle 
and the mass ratio.}
{We conclude that neither the radial velocity curve nor the orbital variation of the
relative donor's contribution to the total flux are affected by irradiation. On the other hand, we do
see emission from the donor star at H${\alpha}$ and HeI $\lambda$5876 which we tentatively attribute to
irradiation effects. In particular, the H${\alpha}$ emission from the companion is clearly asymmetric
and we suggest is produced by irradiation from the hot-spot. Finally, from the velocity
of the HeI $\lambda$5876 spot we constrain the disc opening angle to $\alpha=7^{\circ}-14^{\circ}$.}
      {}


\titlerunning{Irradiation in Cen X-4}

   \maketitle
%

\section{Introduction}

Soft X-Ray Transients (SXRTs), a subclass of Low Mass X-Ray Binaries, are
interacting binary systems which alternate short periods (weeks to months) of
high X-ray luminosity with long (several years) intervals of quiescence. During
quiescence, such transient systems, are very faint in X-rays 
($10^{32} - 10^{33}$ erg s$^{-1}$) and their optical luminosity drops by as much
as 6-7 mag, giving a unique opportunity for the study of companion stars (see Campana et al. 1998 for a
review).

Cen X-4 is a well-known neutron star X-ray transient, from which two
bright X-ray outburst were detected in 1969 (\cite{Co69}) and in 1979 (\cite{Ka80}). 
From then on it remained in the quiescent state. With an evolved 0.3 $\msole$, K3-K7
star (\cite{SNC93}, \cite{To02}, D'Avanzo et al. 2005), Cen X-4 is one of the
brightest SXRTs in quiescence, with $V = 18.7$ and low interstellar absorption ($A_V = 0.3$ mag). A 15.1 hr
orbital period was determined from the sinusoidal variation of the optical light curve (Cowley et al. 1988;
Chevalier et al. 1989; McClintock \& Remillard 1990).
Both the optical spectrum and emission line maps obtained with the Doppler tomography technique (\cite{MH88}) 
show clear evidence of disc emission (\cite{To02}, \cite{PDA05}). 

In general, the effect of heating of the secondary star leads to a change in absorption lines strength, 
with the consequent effect to move the effective centre of the secondary away from the centre of mass of the
star.
This causes phase-dependent distortions of the radial velocity from a sinusoidal fit, leading to 
incorrect determinations of the radial velocity amplitude.

In our previous work on Cen X-4 evidence of irradiation was proposed
in light of a residual emission of H${\alpha}$ and HeI from the companion star. We 
present here a search for evidence of irradiation in Cen X-4, comparing
our observations presented in D'Avanzo et al. (2005) with the simulations 
obtained using the code presented in ~Mu\~noz-Darias, Casares \& Mart\'inez-Pais (2005).

\section{The radial velocity curve of Cen X-4}

\begin{figure}
\epsfig{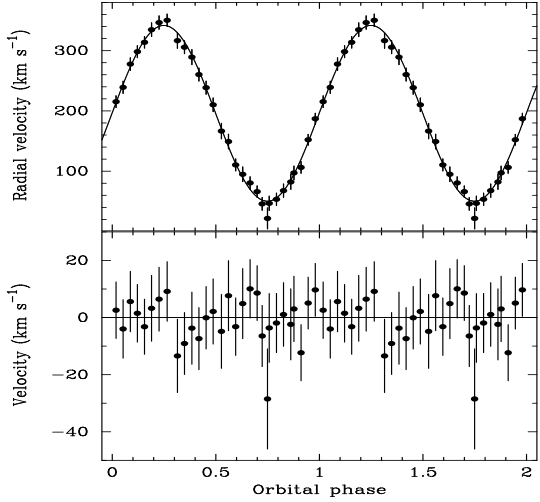}
\vskip 0.0truecm
\caption{Radial velocity curve for Cen X-4 (from~\cite{PDA05}) obtained through cross-correlation of absorption
lines in the range $5930-6750 \AA$ with late$-$type template stars. Two phases are shown for 
clarity. The best sine-wave fit is also shown. Residuals from the best fit are shown in the lower panel. \label{fg:rr_vv}}
\end{figure}

In D'Avanzo et al. (2005) we presented the results of the first extensive and contiguous 
orbital coverage of Cen X-4. In Fig.~\ref{fg:rr_vv}, the radial velocity curve
we obtained for Cen X-4 is shown, together with the residuals to the sinusoidal 
fit. To overcome any spurious effect, we have carefully corrected our spectra for any
telescope flexure, using telluric lines. It is clear from Fig.~\ref{fg:rr_vv} that there
are deviations from the best fit; such deviations are of the order of a few km~s$^{-1}$.

In order to understand if this behaviour could be considered as an indication
of irradiation we have compared our observational result with a simulation
obtained with the code developed by Mu\~noz-Darias, Casares \& Mart\'inez-Pais (2005). 
In this simulation we model the effect of irradiation in radial velocity curves derived 
from the Doppler shift of absorption lines formed by the companion. 
These features are quenched in the heated face of the donor and hence the 
observed $K-$velocity provides only an upper limit to the real $K_2$. This 
deviation roughly depends on the X-ray luminosity, the orbital separation and 
the opening disc angle ($\alpha$) which partly obscures the companion. Here, we have 
only considered the limit case in which the absorption lines are totally 
quenched on the irradiated regions of the donor star which are not shadowed by 
the disc.

\begin{figure}
\epsfig{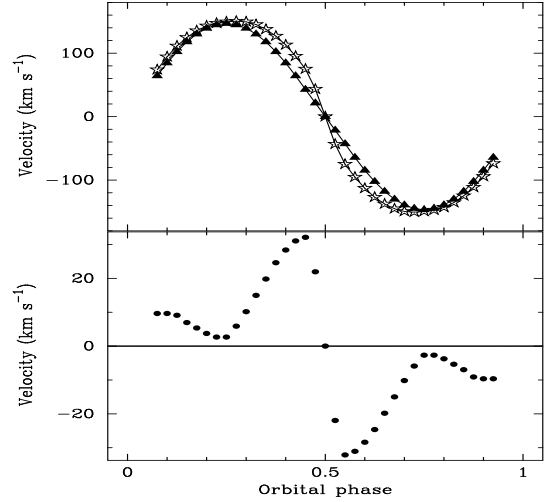}
\vskip 0.0truecm
\caption{Simulated radial velocity curves for Cen X-4 for the irradiated (stars) and non
irradiated (triangles) case. Differences between the two models are shown in the lower panel. \label{fg:rr_vv_sim}}
\end{figure}

We have produced two curves (both showed in Fig.~\ref{fg:rr_vv_sim}), representing the
irradiated and the non-irradiated case. Residuals are shown at the bottom of the
figure. It is immediately clear that the main effect of irradiation
from the neutron star is the rise in the velocity semi-amplitude, as shown. 
Such a behaviour is clearly observed in the
radial velocity curves of systems where the secondary stars are exposed to irradiation 
from the compact object (like, e.g., \cite{FMSJ90a}, \cite{FMSJ90b}).

In light of the results of our simulations, we can conclude that the radial
velocity curve does not show any evidence of irradiation. The expected rise in
the velocity between phases 0.25 and 0.75 (i.e. when the observer sees the inner face of
the companion) is not present. Although deviations from the sinusoidal fit seem to be present 
in our observed curve these are significantly smaller (given the error bars) than predicted by the 
irradiation model, and most important, these deviations are in the opposite direction to
what is expected in an irradiation scenario.

\section{The optical light curve of the companion}

\begin{figure}
\epsfig{file=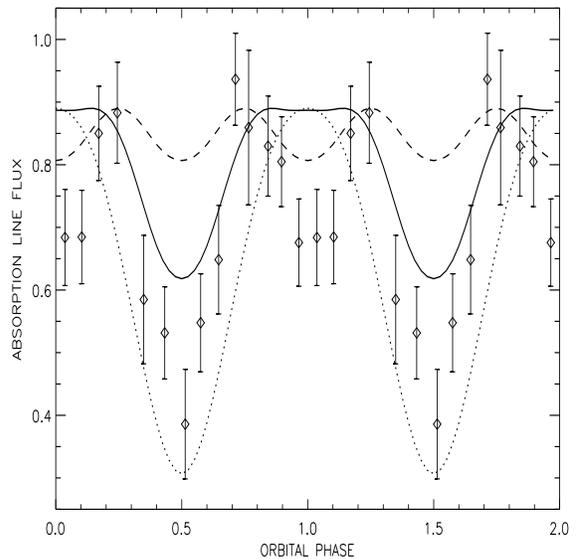, width=8.cm, height=8.0cm, angle=0}
\vskip 0.0truecm
\caption{Variations of the fractional contribution of the companion compared
with synthetic models for $\alpha=0^{\circ}$ (dotted line), 7$^{\circ}$ (solid
line) and 14$^{\circ}$ (dashed line). \label{fg:f_14bins}}
\end{figure}

If the inner side of the secondary star is exposed to irradiation this would also affect
the modulation of the contribution of the companion to the optical luminosity. In D'Avanzo et al. (2005) we
determined the variation of the fractional contribution of light from the secondary (named
factor {\textit{f}}) for 30 phase bins. The result was that {\textit{f}} is not constant at
all the orbital phases, but modulated with two unequal minima at phase 0 (i.e. at superior
conjuction) and 0.5 and two nearly equal maxima at phases 0.25 and 0.75 (i.e. when the
observer sees the sides of the companion with the maximum surface projected area). This 
behaviour is reminiscent of the classical ellipsoidal modulation (see e.g. McClintock \& Remillard 1990;
Shahbaz et al. 1993) caused by the changing visibility of the tidally distorted companion as seen by
the observer. However, we note the unusually large difference in depth between phases 0 and 0.5,
requiring an additional effect like, e.g. the heating of
the inner hemisphere of the companion by X-ray irradiation from the neutron star. In order 
to investigate this scenario we compare the variation of the $f$ factor (coadded into 14 phase bins to
increase the S/N ratio) with a set of synthetic curves obtained with the code developed
by Mu\~noz-Darias, Casares \& Mart\'inez-Pais (2005). This is presented in Fig.~\ref{fg:f_14bins}.

These simulations show the absorption line flux as function of the
orbital phase for three different flaring angles ($\alpha$). It is well established that the temperature of the
outer layers of the companion can be upset by external heating (e. g. Brett \& Smith 1993), 
which removes the vertical temperature gradient of the heated star and quench the absorption lines 
within the irradiated regions. In order to compute our curves we have considered the limit case in 
which absorption lines are totally quenched (i. e. not present) on the irradiated regions of the 
companion (see also Shahbaz et al. 2000). As it is clear in the plot we cannot reconcile the models
and the data for any value of $\alpha$. If the opening angle is low, 
the absorption line is quenched upon the whole irradiated hemisphere of
the companion and hence the absorption line flux is very low at orbital
phases close to 0.5 (dotted line). However, we can not reproduce the lower $f$
factor found for orbital phases around 0. On the other hand, for $\alpha \ge 14^{\circ}$
(dashed line) the accretion disc completely screens out the companion star, the effect of
irradiation disappears  and we obtain the ellipsoidal modulation associated with the changing visibility of
the tidally distorted companion star. 
If we consider the X-ray luminosity of $L_X \sim 10^{32}$ erg s $^{-1}$ reported in Campana et al. (2004) 
and the orbital parameters of the system, we estimate using our binary code
that the effective temperature of the irradiated regions of the donor is only $0-30$
K higher than in the non-irradiated regions. Hence, the donor's atmosphere 
should almost not be affected by heating and we expect an $f$
factor behaviour close to the latter case in which irradiation effects are not present. Therefore, 
as result of our simulations, the irradiation from the inner regions of the accretion disc does not appear to be 
responsible for the orbital behaviour of the $f$ factor.

\section{H${\alpha}$ and HeI emission from the companion}

In D'Avanzo et al. (2005) we presented Doppler maps of Cen X-4 showing clear evidence of residual H${\alpha}$ and HeI
$\lambda$5876 emission associated with the companion star. As opposed to HeI $\lambda$5876, the H${\alpha}$ spot is clearly
offset with respect to the vertical $V_y$ axis. As reported in D'Avanzo et al. (2005) we have computed the
centroids of these spots obtaining:\newline

H$\alpha$: $(V_x, V_y)=(34.1\pm11.8, 97.7\pm11.8)\,{\rm km \ s^{-1}}$

HeI $\lambda$5876: $(V_x, V_y)=(10.3\pm11.8, 122.8\pm11.8)\,{\rm km \ s^{-1}}$\newline

\begin{figure}
\epsfig{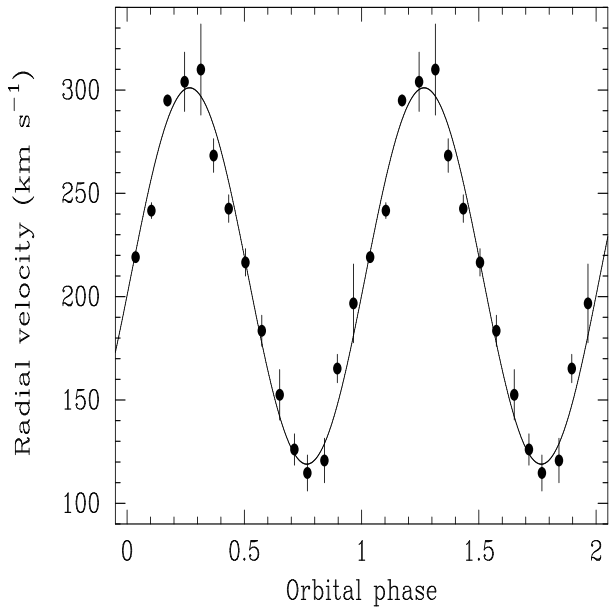}
\vskip 0.0truecm
\caption{Velocity of the H$\alpha$ emission from the companion at different orbital phases. Two
cycles are shown for clarity. 
\label{fg:Halpha_rv}}
\end{figure}

\noindent This clearly demonstrates that the offset of the H${\alpha}$ spot is significant at $\sim
3-{\sigma}$ and it leads phase 0 by 0.05 $\pm$ 0.03 cycles. The S-wave component responsible for the 
H${\alpha}$ spot is clearly seen in the individual spectra. Therefore, in order to reduce the 
uncertainty in the position of
the region of the companion emitting H${\alpha}$ photons we have performed a multi-Gaussian fit
to the H${\alpha}$ profiles, coadded into 30 phase bins. Such operation enabled us to isolate the H${\alpha}$ 
component emitted from the companion for each phase bin, and to determine with good precision its
centroid. As a result we obtained a sinusoid-like velocity curve, shown in
Fig.~\ref{fg:Halpha_rv}, whose semi-amplitude (equal to 91.1 $\pm$ 3.0 km s$^{-1}$) is in agreement 
with the value of $V_y$ obtained from the Doppler map, but it is more precise and gives the velocity 
of the H${\alpha}$ emission region of the companion. Moreover, we also note that 
Fig.~\ref{fg:Halpha_rv} shows also an indication of a phase shift (equal to 0.02 $\pm$ 0.01) which is in 
agreement to the shift in the $V_x$ direction of the H${\alpha}$ spot in the Doppler map.

Unfortunately the lower S/N ratio of the HeI $\lambda$5876 line does not enable us to
do an accurate multi-Gaussian fit to isolate the companion component, so we used the position
given from the Doppler map. In this case no phase/velocity shift is present in the $x$ direction.

In D'Avanzo et al. (2005) we measured an equivalent width for the H${\alpha}$ residual of 4.4 $\pm$ 0.5 \AA~which could
not be explained only invoking chromospheric activity from the companion. Hence, we proposed that the
H${\alpha}$ spot may be triggered by irradiation from the compact star (for details see~\cite{PDA05}). 
This scenario predicts lower excitation lines to be formed at higher Roche lobe latitudes because of variable
photoelectric absorption by the outer disc rim i.e. lower energy photons are more efficiently absorbed and
hence lower excitation lines arise from regions close to the poles. This has been previously seen in the
cataclysmic variable IP Peg (Harlaftis 1999, \cite{MMB00}) and more recently in U Gem
(\cite{UMM06}). This scenario, however, does not
agree with our observed velocities for H${\alpha}$ and HeI emission. Moreover, with a velocity semi$-$amplitude of 
91.1 $\pm$ 3.0 km s$^{-1}$, the H${\alpha}$ residual emission arises from the vicinity of the L$_1$ point 
which must be efficiently shadowed from the UV photons by the disc. This, together with the asymmetric
distribution of the H${\alpha}$ spot over the Roche lobe, points to the hot-spot as a possible source of the 
H${\alpha}$ photoionising radiation.

\section{Discussion}

The discrepancies between the observed and the simulated radial velocity curve enable us to
conclude that our radial velocity curve of Cen X-4 does not contain any evidence of the presence of an 
irradiated companion and that the measured semi-amplitude of the curve must correspond to the 
``real'' K$_2$ velocity of the secondary star. This is also confirmed by new high quality UVES data at
7 km s$^{-1}$ resolution (Casares et al. 2006, in preparation). Therefore, no ``K-correction'' seems 
to be necessary for this system.

The interpretation of the orbital modulation of the optical luminosity contribution of the companion is
more trivial. We note here that the $f$-factor yields not
the lightcurve of the companion star but the relative contribution to the total 
flux. Therefore, instead of neutron star irradiation, a more plausible explanation 
for the deep phase 0.5 minimum
would simply be an increase in the disc brightness around this particular phase
i.e. a non-uniform emissivity  distribution  which might trace azimutal
variations of the disc density.

Neutron star irradiation seems to be ruled out also as responsible for the H${\alpha}$ emission from
the companion. In fact, as we have seen in the previous section, the H${\alpha}$ residual arises from a
region of the companion which should be affected by the shadow of the disc. 

In light of this, hot-spot irradiation seems to be a natural alternative explanation for the intrinsic H${\alpha}$ emission
from the companion, because in such case, no shielding is to be considered. In fact, as we see from our Doppler maps, 
bright anisotropies are present on the accretion disc. We propose that the H${\alpha}$ emission 
concentrates in the trailing
side of the Roche lobe because the leading side is shielded by gas stream.
Hot-spot heating has been previously  invoked to explain 
asymmetric emission patterns observed in several cataclysmic variables (e.g. 
in IX Vel, see Beuermann \& Thomas 1990). Furthermore, evidence for a hot-spot 
or splash point in Cen X-4 was presented by McClintock \& Remillard (1990) 
whereas our HeI Doppler images (Fig. 4 in  D'Avanzo et al. 2005) also show 
evidence of bright spots coincident with the interaction between the outer 
disc rim and the gas stream. Alternatively, asymmetric heating may be amplified 
by the action of circulation currents induced through Coriolis forces (Davey \& 
Smith 1992, Martin \& Davey 1995). A direct consequence of hot-spot (if present) or, in general, anisotropy
irradiation is the shift, in the $V_x$ direction, of the H${\alpha}$ emission region on the companion,
exactly as observed.

\begin{figure}
\epsfig{file=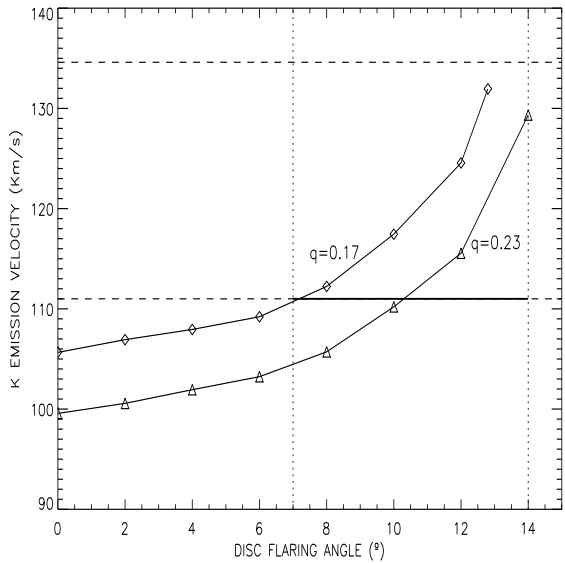, width=8.cm, height=8.0cm, angle=0}
\vskip 0.0truecm
\caption{Determination of the disc flaring angle in Cen X-4. The solid lines represent the result of our model for the
considered mass ratio values ($q \sim 0.17$ and $q \sim 0.23$). The dashed lines represent the value 
$K_{em}=122.8 \pm 11.8$ km s$^{-1}$  measured from or Doppler maps. The possible disc flaring angle range is marked with
a solid line. \label{fg:disc_ap}}
\end{figure}

On the contrary, the higher $V_y$ and the absence of any phase/velocity shift in the $x$
direction, should suggest that the HeI emission region on the companion lies in a region of the Roche 
Lobe that is out from the shadow of the disc and directly exposed to illumination by the neutron
star. We can use this hypothesis to put some strong constraints on the aperture
angle $\alpha$ of the accretion disc. To do this, we have used the K-correction values tabulated in 
Mu\~noz-Darias, Casares \& Mart\'inez-Pais (2005). In this work,  the relation $K_{em}/K_{2}$  
is computed as function of the mass ratio ($q=\frac{M_1}{M_2}$) for 
different values of $\alpha$. Where $K_2$ is the radial velocity 
associated with the mass center of the companion and $K_{em}$ is the 
velocity of an emission line formed on the heated face of the donor. 
Since $q$ and $K_2$ are known for Cen X-4 we have used the observed 
$K_{HeI}$ emission line velocity to constrain the value of $\alpha$. As 
is explained above, the low S/N ratio of this line does not allow us to 
isolate the companion component and hence we have used  the 
$K_{em}=122.8 \pm 11.8$ km s$^{-1}$ obtained from the Doppler map. In 
Fig. 5 we show that using this value of $K_{em}$ and considering $0.17 
\leq q \leq 0.23$ we directly obtain $\alpha \geq 7^{\circ}$. Moreover, 
if we take into account the maximum value of $\alpha$ to the companion 
be irradiated for this $q$, we can further constrain the opening disc 
angle between $7^{\circ} \leq \alpha \leq 14^{\circ}$ (Fig.~\ref{fg:disc_ap}).

\section{Conclusions}

We have examined the H$\alpha$ and HeI $\lambda$5876 bright spots associated 
to the companion star in Cen X-4. The HeI spot is well centered on the line 
joining the two stars and we use its velocity to constrain the disc flaring
angle to be 7-14$^{\circ}$. On the other hand, the H$_{\alpha}$ spot
concentrates around the L1 point and is shifted towards the leading side of the
Roche lobe. We tentatively explain this asymmetry through irradiation from the
hot-spot and shielding by the gas stream. 

We have also searched for evidence of irradiation effects in the radial
velocity curve of the photospheric absorptions, by comparing our data with
model simulations, but found none. This lead us to conclude that the donor's 
metallic lines are not affected by irradiation and hence the orbital parameters
presented in previous papers (Torres et al. 2002, D'Avanzo et al. 2005) are 
safe from bias. 

\begin{acknowledgements}
PDA thanks the Astrophysics Institute of the Canary Islands (IAC) for kind hospitality. SC and PDA acknowledge 
the Italian Space Agency for financial support through the project ASI I/R/023/05. JC and PDA acknowledge support 
from the Spanish Ministry of Science and Technology through the project AYA2002-03570.
\end{acknowledgements}

\end{document}